# Seismic Interpolation Transformer for Consecutively Missing Data: A Case Study in DAS-VSP Data

Ming Cheng, Jun Lin, Xintong Dong, Shaoping Lu, Tie Zhong

*Abstract*—Distributed optical fiber acoustic sensing (DAS) is a rapidly-developed seismic acquisition technology with advantages of low cost, high resolution, high sensitivity, and small interval, etc. Nonetheless, consecutively missing cases often appear in real seismic data acquired by DAS system due to some factors, including optical fiber damage and inferior coupling between cable and well. Recently, some deep-learning seismic interpolation methods based on convolutional neural network (CNN) have shown impressive performance in regular and random missing cases but still remain the consecutively missing case as a challenging task. The main reason is that the weight sharing makes it difficult for CNN to capture enough comprehensive features. In this paper, we propose a transformer-based interpolation method, called seismic interpolation transformer (SIT), to deal with the consecutively missing case. This proposed SIT is an encoder-decoder structure connected by some U-shaped swin-transformer blocks. In encoder and decoder part, the multi-head self-attention (MSA) mechanism is used to capture global features which is essential for the reconstruction of consecutively missing traces. The U-shaped swin-transformer blocks are utilized to perform feature extraction operations on feature maps with different resolutions. Moreover, we combine the loss based on structural similarity index (SSIM) and $L_1$ norm to propose a novel loss function for SIT. In experiments, this proposed SIT outperforms U-Net and swin-transformer. Moreover, ablation studies also demonstrate the advantages of new network architecture and loss function.

## I. INTRODUCTION

DISTRIBUTED optical fiber acoustic sensing (DAS) is a novel acquisition technique for seismic data. It uses the light scattering response of laser pulses to record the strain changes caused by seismic events. Recently, DAS arrays have been extensively utilized to acquire downhole seismic data due to their inherent advantages such as high sensitivity, high acquisition density, and expanded dynamic range. However, some facts, such as the fiber cable breakage, obstacle, and inferior coupling between cable and well, lead to the phenomenon of missing traces in vertical seismic profile (VSP) data acquired by DAS system, seriously affecting the quality of DAS-VSP data. The consecutively missing case, or called big gap, is relatively challenging. Also, the reconstruction of this consecutively missing case is compounded by various DAS background noise with strong energy. Thus, it is necessary to explore an effective method to interpolate the consecutively missing traces in DAS-VSP data and thus provide high-quality data for the following imaging [1], [2] and interpretation [3], [4] tasks.

Traditional seismic interpolation methods can be roughly divided into four categories: wave-equation-based methods, low-rank-based methods, predictive filtering-based methods, and sparse-based methods. Specifically, wave-equation-based methods [5], [6] employ the wave equation to estimate missing seismic traces by utilizing neighboring traces as constraints for accurate reconstruction. However, the reconstruction performance of such methods heavily relies on the accuracy of the initial velocity model. Low-rank-based methods [7], [8] such as Cadzow filtering [9], [10], singular spectrum analysis (SSA) [11], [12], and principle component analysis (PCA) [13], assume that the incomplete matrix (i.e. missing traces seismic record) has a high-rank structure and interpolation task can be accomplished by reducing its rank. Nonetheless, how to select the appropriate parameters is a tough problem for this kind of method. Predictive-filtering-based methods [14]-[16] achieve seismic interpolation by selecting suitable filer parameters to misfit the difference between the input data and the desired data. Typical methods include f-x predictive filtering [17], [18], τ-p predictive filtering [19], and Kalman filtering [20]. Notably, these methods are sensitive to the selection of crucial parameters, and the inappropriate parameter setting always brings degenerate interpolation results. Sparse-based methods [21], [22] are another kind of mainstream interpolation methods, mainly include wavelet [23], [24], seislet [25], [26], curvelet [27], [28], shearlet [29], radon [30], [31], and dictionary learning [32], [33]. These methods reconstruct missing traces by representing the seismic data with a sparse set of basis functions and solving an optimization problem to find the sparse coefficients. Nonetheless, this kind of interpolation method is sensitive to the selection of sparse representation

This work was financially supported in part by the National Natural Science Foundation of China under Grant 42204114, and in part by the Graduate Innovation Fund of Jilin University under Grant 2024CX100(Corresponding author: Xintong Dong).

Ming Cheng, Jun Lin, and Xintong Dong are with the College of Instrumentation and Electrical Engineering, Jilin University, Changchun, Jilin 130026, China. (e-mail: chengming22@mails.jlu.edu.cn; lin_jun@jlu.edu.cn; 18186829038@163.com).

Shaoping Lu is with the School of Earth Sciences and Engineering, Sun Yat-Sen University, Guangzhou, Guangdong 510275, China, also with Southern Marine Science and Engineering Guangdong Laboratory (Zhuhai), Zhuhai, Guangdong 519000, China, and also with Guangdong Provincial Key Lab of Geodynamics and Geohazards, Sun Yat-sen University, Guangzhou, Guangdong 510275, China (e-mail: lushaoping@mail.sysu.edu.cn).

Tie Zhong is with the Key Laboratory of Modern Power System Simulation and Control and Renewable Energy Technology (Ministry of Education), Jilin 132012, China, and also with the Department of Communication Engineering, College of Electric Engineering, Northeast Electric Power University, Jilin 132012, China (e-mail: 519647817@qq.com).

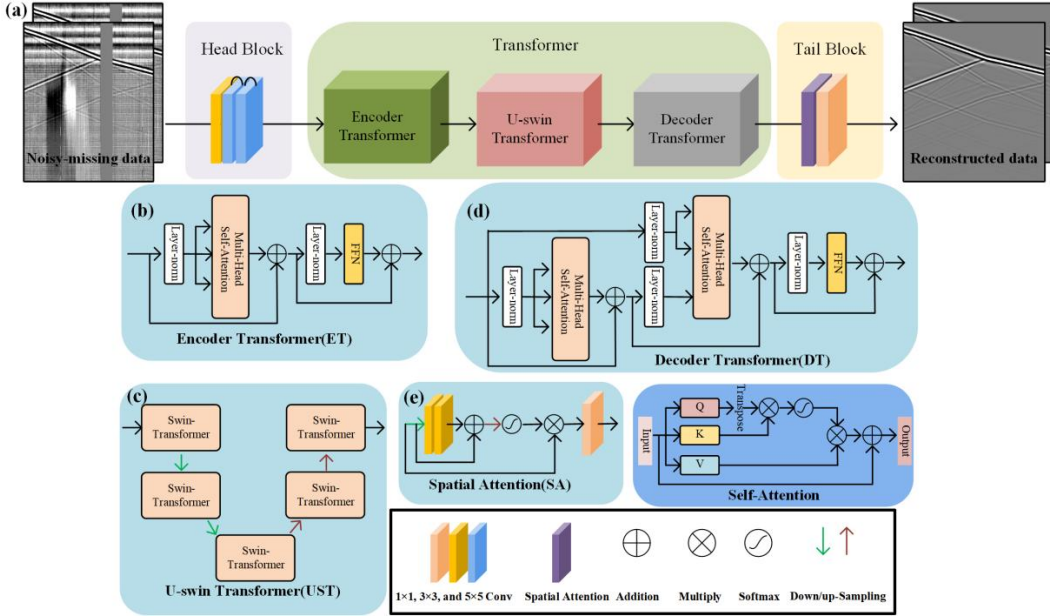

**Fig. 1.** Architecture of Seismic Interpolation Transformer (SIT).

and dictionary, which may impact the accuracy of the reconstruction results, particularly in the presence of complex subsurface structures. Although these existing methods have shown promising performance in seismic data interpolation, they have two limitations: For one thing, these methods are theory-driven, and their performance heavily depends on mathematical or physical properties. For another, these methodologies are primarily tailored for regular or random instances of missing data, and show limited effects when confronting seismic data with consecutive missing traces. It means that the traditional methods often have their downsides and provide unexpected reconstruction results in some conditions. Therefore, the interpolation of DAS seismic data with large-scale consecutive missing traces has become a bottleneck that urgent to be overcome.

Recently, deep-learning (DL) has been proven to be one of the most promising tools for seismic data processing. DL can establish some non-linear relationship between input and output and thus automatically accomplish some tasks, such as denoising [34], [35], interpolation [36], [37], velocity inversion[38], arrival time picking [39]-[41], and fault detection [42], [43]. For interpolation, Liu et al. [44] construct an innovative framework combined with residual neural network and wavelet transform; which provides promising reconstruction performance for seismic data interpolation. Wang et al. [45] propose an algorithm based on residual dense network (RDN) for noise suppression and interpolation of pre-stack seismic data. Also, the network achieves remarkable performance in both regular and irregular conditions. Yoon et al. [46] adopt the generative adversarial network (GAN) to accomplish the interpolation task via an unsupervised manner. Although these DL-based interpolation methods show good performance in regularly and randomly missing cases, how to reconstruct the consecutively missing traces is still a challenging task. Meanwhile, these DL-based interpolation methods mainly adopt CNN-based frameworks, such as U-Net, Res-Net and GAN. Notably, CNN extracts features by performing small convolutional kernel on data, so it is difficult to capture global information, which significantly hinders the reconstruction of consecutively missing traces. On this basis, some experts utilize attention mechanism to help CNNs to capture more global features and then enhance their performance in consecutively missing case. Yu et al. [47] comprise a CNN architecture with attention mechanism for consecutively missing seismic data reconstruction; this proposed A-Net can effectively address the reconstruction of missing traces with big gaps and ensure continuity in recovering weak signals. Liu et al. [48] proposed an attention-guided U-Net with a hybrid loss to enhance the reconstruction capability, so as to achieve the reconstruction of consecutive missing traces in seismic data. Li et al. [49] introduce the coordinate attention mechanism into the U-Net architecture, which can reconstruct consecutively missing seismic data at different stages of U-Net. Although these methods have gained remarkable performance, their reconstruction results are likely to degrade when influenced by intense background noise. Compared with conventional seismic data received by electronic geophones, the big gap is more common in DAS-VSP data. In addition, a variety of DAS background noise with strong energy further aggravates the difficulty of this reconstruction task. Therefore, the reconstruction of consecutively missing traces in DAS-VSP data is a special task that needs to explore a targeted and effective method.

Transformer, an emerging deep learning-based architecture, has garnered the attention of researchers in the area of seismic data processing. Compared with CNN, the transformer exhibits enhanced capability in capturing comprehensive features through a large-scale acquisition provided by multi-head self-attention. [50] and [51] have proved its powerful capability of global feature extraction. Therefore, we propose the seismic interpolation transformer (SIT), a transformer-based architecture for DAS-VSP data interpolation, which can effectively recover consecutively missing traces. Specifically, SIT is a transformer-based network with an encoder-decoder architecture, comprises five key components: the head block (HB), the encoder transformer (ET), the U-shaped swin-transformer (UST), the decoder transformer (DT), and the tail block (TB). The HB is utilized to initially extract the features from the training data set. Subsequently, we use transformer-based architecture including ET, UST, and DT to extract global features. Among them, a UST with multi-cascade structure excels at capturing different resolutions and leverages the cross-window information to enhance feature interaction. By combining the feature extraction capabilities of the three transformer-based blocks, both temporal and spatial dimension information can be sufficiently extracted. Finally, a TB is employed at the end of SIT to integrate the obtained features and prepare to output final reconstruction results. In addition, inspired by the structural similarity index measure (SSIM) and $L_1$ norm, we design a novel loss function called SSIM_$L_1$ to optimize the parameters of the interpolation model during the training process. Moreover, a high quality training data set is also constructed to boost the effectiveness of the model. Experiments on synthetic and field seismic data have demonstrated the superiority of SIT for the interpolation of consecutively missing traces in DAS seismic data.

## II. NETWORK ARCHITECTURE

In this study, we propose a transformer-based method, called SIT, to interpolate the consecutively missing traces in DAS seismic data. As shown in Fig. 1, the proposed SIT is composed of five major components: head block (HB), encoder transformer (ET), U-swin-transformer (UST), decoder transformer (ET), and tail block (TB). Firstly, the HB, consisting of three convolutional layers with different kernel sizes, is employed to extract some shallow features. Moreover, the encoder transformer and decoder transformer (including ET and DT) are connected by five USTs. The ET and DT based on multi-head self-attention mechanism (MSA) can capture the global features in seismic data. Meanwhile, the UST is performed on feature maps with different resolutions. Finally, we use the TB, composed by spatial attention block and 1×1 convolutional layer, to enhance and refine the extracted features.

### A. Head block (HB)

Here, HB is composed of a 3×3 convolutional layer and two followed 5×5 convolutional layers, which are all activated by the leaky ReLU. Also, we add a residual connection to each 5×5 convolutional layer to avoid the over fitting issue. Moreover, the size of input and output is set to H×W (height×width) and the channel is expanded from 1 to 64. In this study, we set H and W to 64. Notably, the main function of HB is to extract the shallow features.

### B. Transformer-based feature extraction structures

The output of HB is propagated into a series of transformer-based structure, including ET, DT, and UST, to extract global and multi-resolution features.

1) Encoder Transformer (ET) block

Fig. 1(b) gives the architecture of ET block. In ET block, we first employ a layer norm (LN) to process the output features of HB. Subsequently, an MSA with six heads is utilized to capture long-range dependencies in the seismic data. The obtained information is then input into LN and a followed feed-forward network (FFN) is utilized to further enhance the obtained features. Additionally, residual connections are also employed to facilitate the interaction of informative features. Specifically, the detailed operation of ET can be concluded as follows:

The features at stage k-1 in ET operation can be expressed as
$$y'_k = MSA(LN(y_{k-1})) + y_{k-1} \quad (1)$$
$$y''_k = FFN(LN(y'_k)) + y'_k \quad (2)$$
where $y_{k-1}$ is the input at k-1 stage, and $y''_k$ is the final output of the ET block. Meanwhile, the MSA, FFN, and LN, represent the processing of aforementioned network components, respectively.

2) U-shaped swin-transformer (UST) block

The output of ET block is transmitted to the UST block, whose architecture is shown in Fig. 1(c), aiming to obtain more comprehensive information with different resolutions. The UST block consists of five swin-transformer (ST) blocks and two pairs of down-sampling and up-sampling operations. Specifically, the ST follows the structure of classical transformer, which is primarily composed of shifted window multi-head self-attention (SW-MSA) and FFN. Among ST, the SW-MSA plays a crucial role by allowing the model to attend to a limited region around each position in the data, so as to capture long-range dependencies and meet the requirements of texture feature extraction. Building upon this foundation, we arrange five ST blocks following the shape of classical U-Net architecture. Initially, two STs combined with downsampling operations are used to capture the shallow features in the low-resolution of seismic data. During this process, we shrink the size of features to a quarter and quadruple the number of original channels to avoid feature loss. Subsequently, an additional ST is employed to extract contour features, and two STs combined with upsampling operations are utilized to recover the feature maps and the channels to their original size. By using UST, both the temporal and spatial dimension features can be highlighted and enhanced.

3) Decoder Transformer (DT) block

Similar to the architecture of ET block, the DT block, as shown in Fig. 1(d), first refine the input features by using an LN and an MSA. The corresponding output is added to the original input of DT and then transfers into the Q path of the second MSA after an LN layer. To ensure the accuracy of the captured features, we also transfer the input of DT after an LN layer to the K and V paths of the second MSA. Subsequently, an LN layer is used to preliminarily extract the potential features exist in the outputs of the first and second MSA. On this basis, FFN is employed to further enhance the extracted features with the assistance of the residual connection.

4) Principle of multi-head self-attention mechanism (MSA)

In each transformer-based feature extraction block, MSA is used as a main network component extract the long-range information. Denoting the input data as $X \in R^{H \times W \times d}$, the basis procedure of MSA can be concluded as follows:

For each head $i$, we apply three learned linear transformations to the input data:
$$Q = X \cdot W_{Q_i}$$
$$K = X \cdot W_{K_i} \quad (3)$$
$$V = X \cdot W_{V_i}$$

where $W_{Q_i}$, $W_{K_i}$, $W_{V_i} \in R^{d \times d_h}$ denote the learned weight matrices for the Q, K, and V transformations, respectively. Moreover, $d$ and $d_h$ represent the dimension channel and hidden space for each head. On this basis, we can compute the scaled dot-product attention scores for each head:

$$MSA_i(Q_i, K_i, V_i) = \text{softmax}(\frac{Q_i \cdot K_i^T}{\sqrt{d_h}}) \quad (4)$$

where softmax is a non-linear activation function commonly used in DL. The scaling factor $\sqrt{d_h}$ is used to prevent the dot product from having a over large value, which can lead to a very small gradients during the training process. Finally, we concatenate the outputs from all heads and project them back to the original dimension:

$$Concat_i = Concatenate(MSA_1, \ldots MSA_h) \quad (5)$$
$$MSA_{output} = Concat_i \cdot W_O \quad (6)$$

Here, $W_O \in R^{h \cdot d_h \times d}$ is another learned weight matrix for the output projection, and $h$ is the number of attention heads.

*C. Tail block*

In this study, we design a TB to enhance the desired features and extract relevant information to reinforce the feature representation. Specifically, the TB consists of a spatial attention mechanism (SA) block and a 1×1 convolutional layer. Generally, the SA block, as shown in Fig. 1(e), is designed to emphasize informative features, while the convolution layer is used to prepare the final output. For the architecture of SA block, we first downsample the input to half of its original size, and double the channel numbers to avoid feature loss. Subsequently, two 3×3 convolutional layers with the residual connection are utilized to further extract the contour information. Then, we use up-sampling operation to recover the size of features to original. On this basis, softmax function is utilized to obtain the distribution of the effective features, which is further used as the weights to highlight and enhance the primary features. At the end of TB, the optimized information is transferred to a 1×1 convolutional layer, to prepare to output the final reconstruction result.

*D. Loss function*

In the training process of the DL-based methods, the parameters of model are optimized by loss function. Specifically, the loss function usually optimizes the parameters by minimizing the difference between the predictive (reconstruction data) and the target data (complete record). An appropriate loss function can accelerate the training process, and make network convergence better. The mean square error (MSE), a widely utilized loss function, is computed at the pixel level and solely quantifies the difference between two pixels. Consequently, the MSE only considers the information in the temporal domain and ignores the characteristics in the spatial domain, which potentially creates unsatisfactory reconstruction results. To address this issue, we combine the traditional $L_1$ norm and SSIM and propose a novel loss function called $L_1\_SSIM$. As we have known, SSIM is a widely used metrics for measuring the similarity between different images. Specifically, SSIM can capture a large range of textural information by considering the potential features exist in luminance, contrast, and structure. Specifically, the calculation of SSIM, used in generating the loss function, is defined as below:

$$SSIM(x, y) = \frac{(2\mu_x\mu_y + c_1)(2\sigma_{xy} + c_2)}{(\mu_x^2 + \mu_y^2 + c_1)(\sigma_x^2 + \sigma_y^2 + c_2)} \quad (7)$$

where x and y are the input and output data, $\mu_x$ and $\mu_y$ represent the means of x and y, while $\sigma_x$, $\sigma_y$, and $\sigma_{xy}$ denotes their standard deviations and covariance, respectively. Meanwhile, $c_1$ and $c_2$ are constant terms, which aims to make the equation always have a convergent numerical results.

Considering the aforementioned limitations of MSE, $L_1$ norm is always more appropriate to construct the loss function. Therefore, the proposed $L_1\_SSIM$ loss function is defined:

$$L_1\_SSIM = \frac{1}{2N}|SSIM(x, y) - 1| \quad (8)$$

where || represents the calculation of $L_1$ norm, and N is the number of input. Notably, the value of SSIM is 1 when the input data equals to the output data.

## III. INTERPOLATION THEORY AND TRAINING PROCESS

*A. Interpolating Theory and Training Strategy*

In general, the acquired seismic record, suffered from intense noise background noise and consecutively missing traces can be denoted as follow:

$$y = A(e + n) \quad (9)$$

where y is the seismic record, e and n are the target data (i.e. complete data) and background noise. A is a matrix composed of 0, 1 column vectors with the same size as y, to simulate the acquisition process of instruments. Through the training process, a nonlinear relationship between the input data y and target data e can be established, and the reconstruction result can be represented as

$$\tilde{y} = SIT(y, \theta) \quad (10)$$

where θ is the training parameters. On this basis, we optimize the nonlinear relationship by minimizing the loss function shown in equation 11, and the final reconstruction results $e_{rec}$ can be obtained as equation 12.

$$L = \frac{1}{2N}\sum_{i=1}^{N}|SSIM(SIT(y_i, \theta), e_i) - 1| \quad (11)$$
$$e_{rec} = SIT((y_i, \theta_{opt})) \quad (12)$$

where $e_i$ and $y_i$ represent the patches of label and input data in training dataset, while $\theta_{opt}$ is the optimal parameters sets.

*B. Construction of Training Dataset*

The DL-based methods often leverage the feature learning to circumvent the intricate process of manual parameter optimization. In this study, SIT is trained in a supervised manner. Therefore, the quality of training dataset plays an essential role in the generation of optimal reconstruction model. Generally, the training set is composed of a complete dataset and its paired missing dataset. For the complete dataset, forward modeling method method is applied and 60 stratigraphic models with varying velocity distributions are constructed. From this basis, these stratigraphic models are excited by Ricker wavelets with different fundamental frequencies, and the propagation of wave fields is simulated by wave equations, solving by finite difference method. The detailed modeling parameters are listed in Table I. Finally, 60 complete and noise-free synthetic seismic records are generated, and 178605 patches, dividing by a 64×64 sliding window, are obtained to compose the complete dataset. For the missing dataset, we consecutively remove 40-50 traces from each complete patch to simulate the effects of consecutively traces. Moreover, noise data from field seismic records is also divided into 64×64 patches and added to the aforementioned missing patches to imitate the influence of background noise. Notably, the noise data is collected from the pre-arrival areas of field DAS-VSP data, which means no reflection signals are involved in the missing patches. In the training process, the complete (label data) and missing patches are fed into the SIT. Fig.2 gives some typical patches derived from the complete and missing dataset.

TABLE I.
PHYSICAL PARAMETER SETTING OF FORWARD MODELS

| Parameters | Specifications |
|---|---|
| Seismic Wavelet | Ricker |
| Central Frequency | 10-80Hz |
| Wave Velocity | 1000-4500m/s |
| Density | 1272-2500kg/m³ |
| Sampling Interval | 1m |
| Sampling Frequency | 2500Hz |
| Well Depth | 500-5000m |

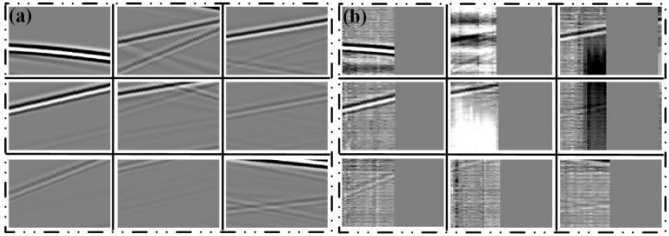

**Fig. 2.** Samples of the training dataset. (a) nine complete signal patches; (b) nine missing data patches.

*C. Indicators*

In this study, we employ the reconstruction assessment factor Q and SSIM (in equation 7) to quantitatively evaluate the performance of different reconstruction methods.
Here, Q value is defined as:

$$Q = 10 \log_{10} \frac{\|R^{clean}\|^2}{\|R^{out} - R^{clean}\|^2} \qquad (13)$$

where $\|\cdot\|$ denotes Frobenius norm, $R^{clean}$ and $R^{out}$ represent the clean complete data and the reconstruction result. Specifically, the larger value of Q and SSIM indicate promising reconstruction capability. In actual seismic data processing, a reconstruction result with a Q value larger than 7dB means the missing traces can be viewed to be well restored.

## IV. RESULTS

*A. Synthetic example*

In this section, we conduct numerical experiments to investigate the reconstruction performance of SIT. Fig. 3 displays a synthetic stratigraphic model, which consists of four horizontal formations with different propagation velocities. We employ a Ricker wavelet with a fundamental frequency of 50Hz to excite the stratigraphic model and obtain the synthetic

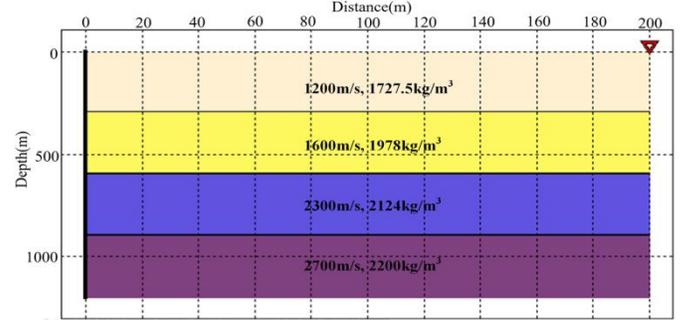

**Fig. 3.** Forward model of the synthetic seismic data.

TABLE II.
NETWORK PARAMETERS OF U-NET AND SWIN-TRANSFORMER

| Hyperparameters | U-Net | Swin-transformer |
|---|---|---|
| Optimizer | ADAM | ADAM |
| Patch size | 64×64 | 64×64 |
| Batch size | 64 | 16 |
| Epoch number | 50 | 100 |
| Learning rate range | [10⁻³,10⁻⁵] | [10⁻³,10⁻⁵] |
| Number of heads | - | 6 |
| Window size | - | 16×16 |

DAS-VSP record, as depicted in Fig. 4(a). Generally, the synthetic data is noise-free, and composed of 500 traces with trace interval and sampling frequency of 1m and 2500Hz, respectively. Noise data (Fig. 4(b)) collected from field DAS record are added to the clean synthetic record, and then 50 traces are consecutively removed from the noisy record to generate incomplete record, as depicted in Fig. 4(c). Notably, the existence of consecutive missing traces significantly corrupts the integrity of the reflection events, while the existence of the intense background noise makes the situation become worse.

To effective recover the missing traces, SIT and two popular DL frameworks, including U-Net [52] and swin-transformer are applied to process the incomplete record. Specifically, U-Net is composed of four scales, and the network depth is 23. For the swin-transformer architecture, we stack nine swin-transformer blocks in a U-shape, aiming to balance the reconstruction capability and computational burden. Detailed parameters of U-Net and swin-transformer are listed in Table II. To ensure a fair comparison, all DL-based methods have been optimized to their best performance, using the same training dataset and targeted hyper-settings.

Fig. 5 gives the reconstruction results obtained by SIT and other two competing methods on the incomplete synthetic DAS record depicted in Fig. 4(c). It is worth noting that all these methods can recover the consecutive missing data to a certain extent, even under the influence of intense background noise. However, the results obtained by U-Net (Fig. 5(a)) and swin-transformer (Fig. 5(b)) are all affected by remained noise and signal leakage, such as the areas marked by the red and green arrows. Additionally, an interesting area in Fig. 5(b) is also highlighted by the red dash ellipses, and some fake events are observed, indicating the reconstruction performance of swin-transformer needs further improvement. In contrast, the result of SIT (Fig. 5(c)) exhibits a conspicuous absence of residual noise or unsatisfactory restored events. All these results demonstrate that SIT has a better reconstruction capability for the incomplete seismic records suffering from consecutive missing traces and strong background noise. Meanwhile, the defects of the reconstruction results obtained by U-Net and swin-transformer also show the inadequacy of these two frameworks. As discussed above, U-Net is a typical CNN-based method, which is lack of extraction ability of global features, leading to shortcomings in the reconstruction of consecutively missing traces. On the contrary, swin-transformer mainly concentrates on the extraction of global features and neglects the global features, bringing negative impacts on noise attenuation. Fig. 5(d) to (f) gives the difference plots between the reconstruction results and complete record (see Fig. 4(a)). Notably, the discrepancy record of SIT in Fig. (c) represents minimal signal leakage compared to the competing frameworks, also approving its promising performance in the reconstruction of consecutive missing data.

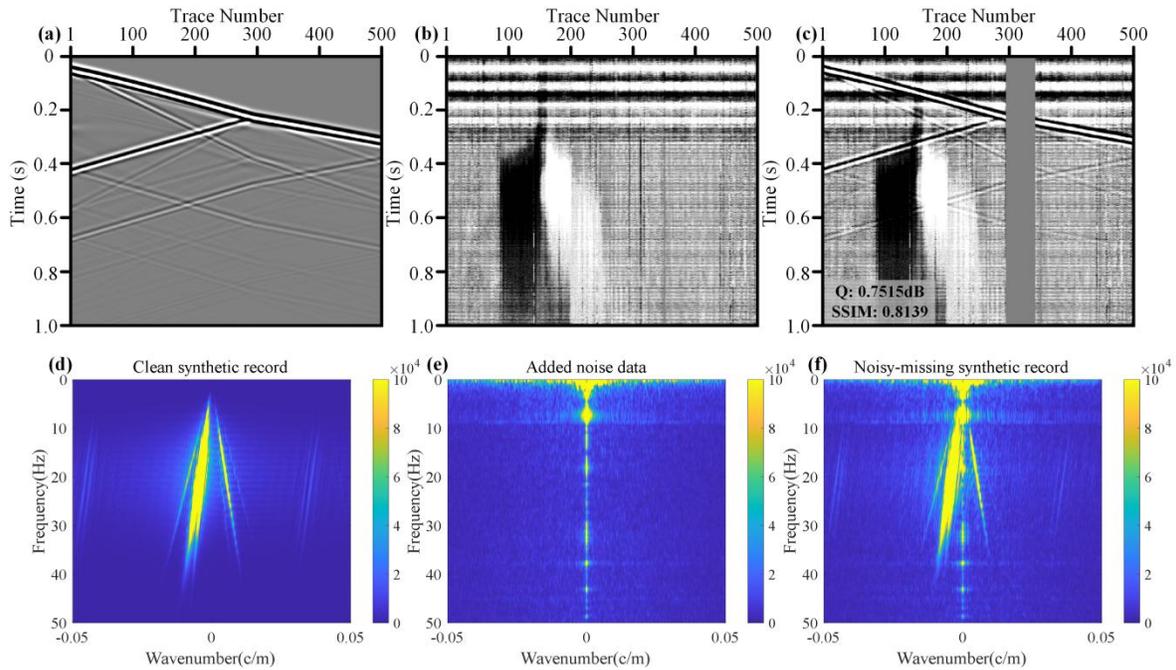

**Fig. 4.** Analyzed records and their f-k spectra. (a) Complete synthetic record; (b) Pre-arrival noise data; (c) Synthetic incomplete record; (d)-(f) F-K spectrum for the corresponding records.

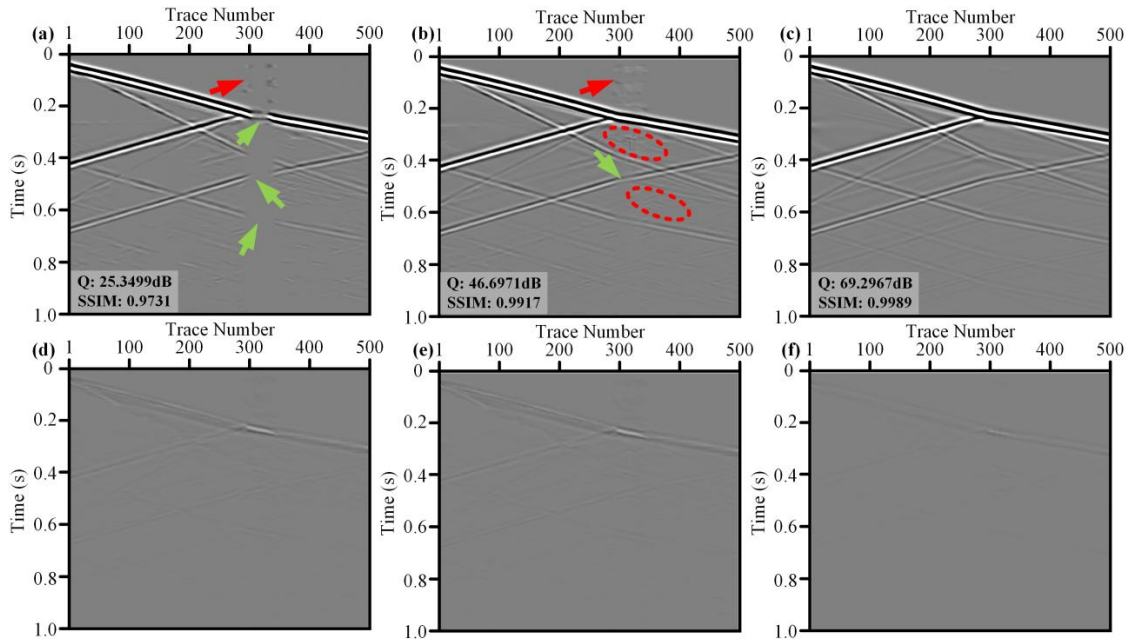

**Fig. 5.** Comparisons of the synthetic reconstruction results and the corresponding difference record. (a)-(c) The reconstruction result of U-Net, swin-transformer, and SIT, respectively; (d)-(f) The difference record between the complete record and the corresponding results.

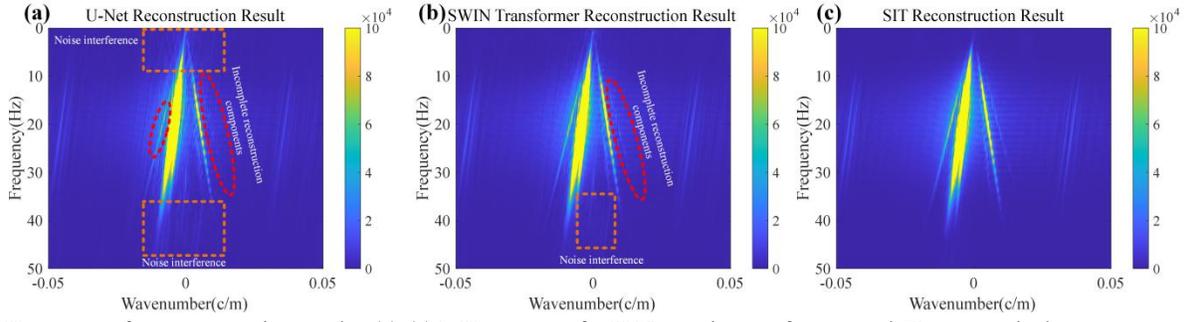

**Fig. 6.** F-K spectrum for reconstruction results. (a)-(c) F-K spectrum for U-Net, swin-transformer, and SIT, respectively.

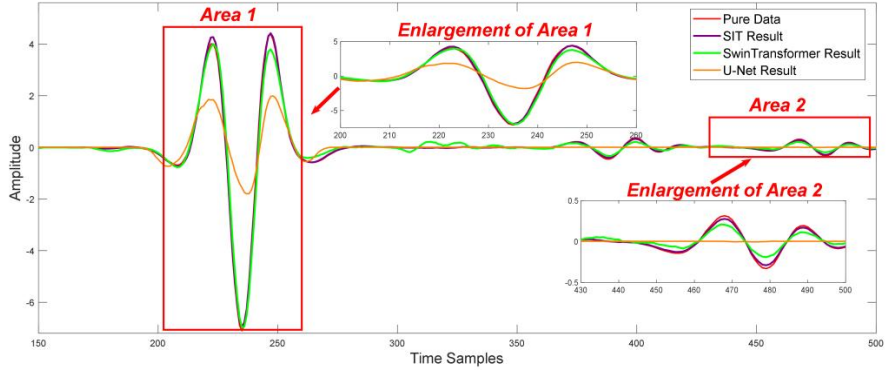

**Fig. 7.** Synthetic trace example, extracted from Fig. 5 with the trace number of 329.

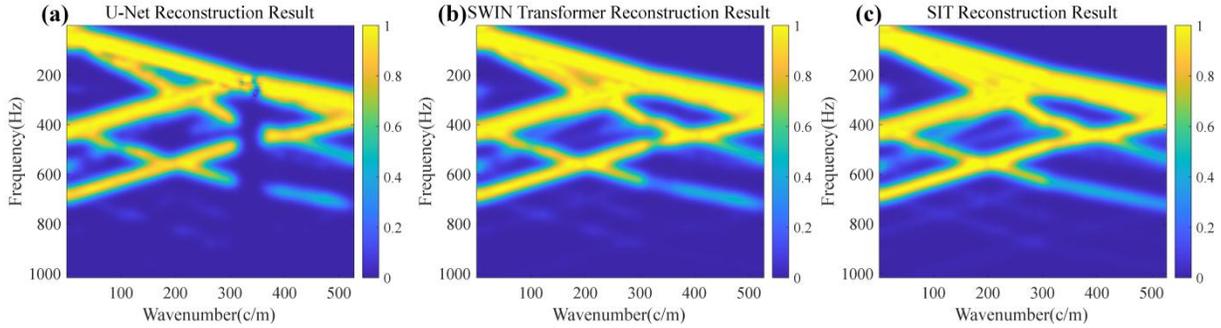

**Fig. 8.** Comparisons of local similarity results. (a)-(d) Results of U-Net, swin-transformer, and SIT, respectively.

For a further comparison, the spectral properties of reconstruction results are also investigated, and Fig. 6 gives the F-K spectrum of corresponding results. As shown in Fig. 6(a), the recovered signals of U-Net are still perform poor continuity (indicated by red dash ellipses) and affected by the residual noise (indicated by orange dashed rectangles). Meanwhile, F-K spectrum of swin-transformer (Fig. 6(b)) shows superiority over the U-Net, owing to its advantages in the extraction of global features. However, obvious noise component can also be noticed, indicating its limited noise attenuation ability. In contrast, SIT (Fig. 6(c)), exhibits advantages over competing approaches in the reconstruction of consecutive missing traces. As discussed above, all these promising results of SIT are attributed to its excellent performance in extracting large-scale and multi-resolution features hidden in seismic data. To check the performance in the preservation of signal amplitude, a typical missing trace (no.329th) is selected for further analysis. Fig. 7 plots the corresponding reconstruction results obtained by different methods. Among the DL-based frameworks, only SIT (the purple line) can properly restore the desired signals, without significantly loss on the signal amplitude. Meanwhile, we also enlarge two areas of interest for a detailed comparison. Similar results are obtained that the restored signal by SIT has the most similar properties to the complete data, indicating its capability in the restoration of consecutive missing data.

Local similarity [53] is another effective tool to evaluate the performance of the reconstruction methods. For clarity, the detailed calculation process of local similarity is given in the APPENDIX. Fig. 8 gives the corresponding results between complete data (Fig. 4(a)) and different reconstruction records. Generally, the bright color means large similarity index and close properties with the target data. Although all DL frameworks show remarkable capability, the recovered events of U-Net (Fig. 8(a)) and swin-transformer (Fig. 8(b)) are corrupted, indicated by the small value of similarity index. Conversely, SIT, as shown in Fig. 8(c), performs better than the competing methods, and areas rich in reflection events are clearly to identified, with bright color and large metrics. In summary, synthetic examples demonstrate that SIT is competent in seismic data reconstruction, even for the DAS-VSP data suffering from consecutive missing traces.

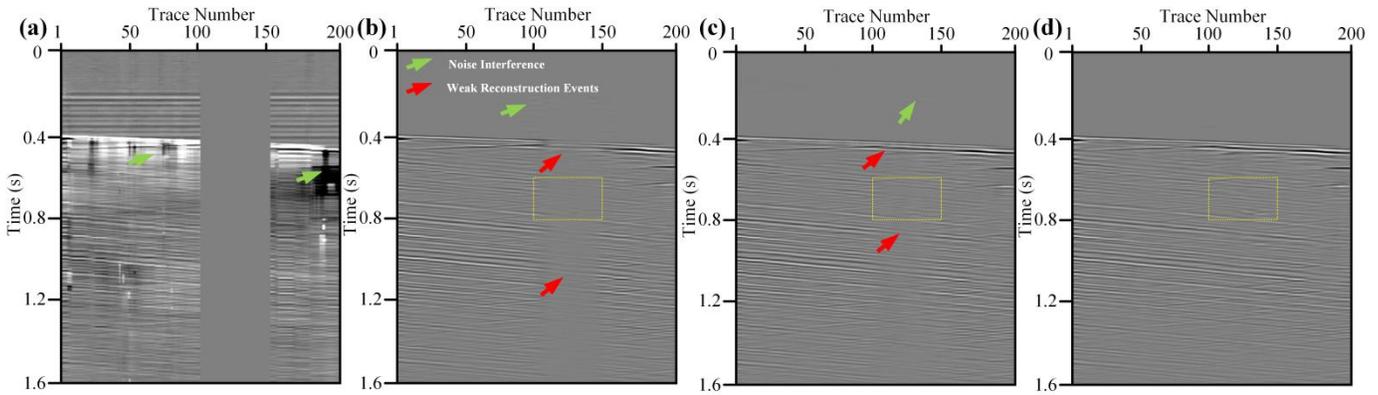

**Fig. 9** Comparisons of the field record 1 reconstruction results. (a) Field incomplete record 1. (b)-(d) The reconstruction result of U-Net, swin-transformer, and SIT, respectively.

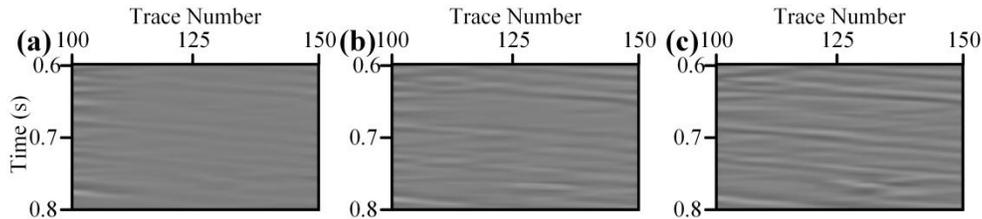

**Fig. 10** Local comparisons of different methods. (a)-(c) Reconstruction result of U-Net, swin-transformer, and SIT, respectively.

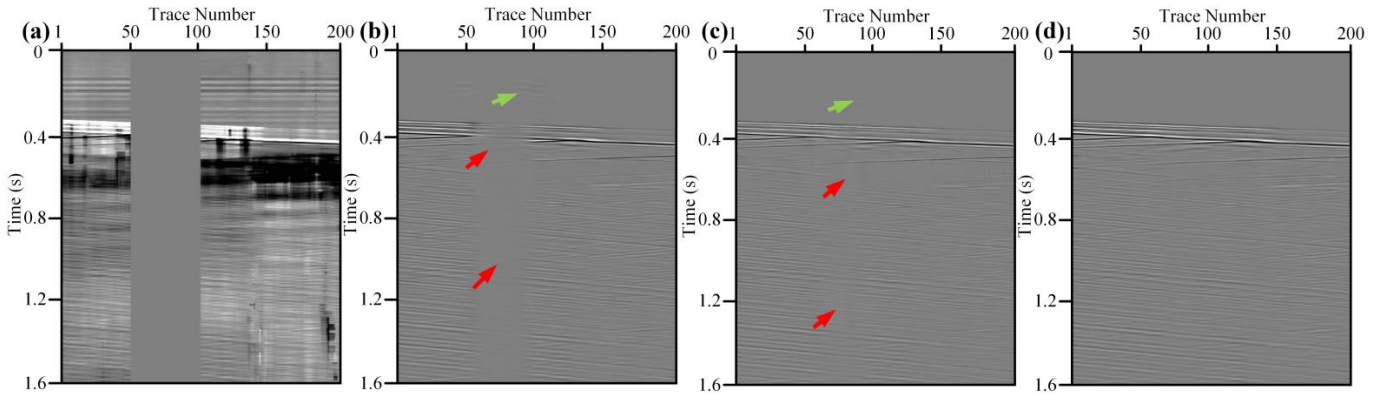

**Fig. 11** Comparisons of the field record 2 reconstruction results. (a) Field incomplete record 2. (b)-(d) The reconstruction result of U-Net, swin-transformer, and SIT, respectively.

*B. Field Example*

In this section, a field DAS-VSP record, shown in Fig. 9(a), is processed by SIT and other two competing methods. Generally, the DAS data is composed of 200 traces, and 50 traces are consecutively missed. Additionally, time interval and sampling frequency of the DAS array are 1m and 2500Hz, respectively. Apart from the consecutive missing traces, intense noise (indicated by green arrows) also contributes to the low quality of the DAS record. Unfortunately, it is impossible to perfectly tell the clean reflection signals from the field seismic data. Therefore, we cannot construct complete training dataset for the field data.

Here, we do not retrain the network, and still use the trained models for synthetic data to process the field record. The similarity between the synthetic and field data, such as similar acquisition parameters and geological models, and the generalization ability of the trained models can ensure the effectiveness of the aforementioned models to a certain degree. By comparing the results in Fig. 9(b) to (d), we can get the point safely that SIT has the most promising reconstruction ability. For one thing, the results obtained by U-Net (Fig. 9(b)) still have obvious remaining noise (marked by green arrows) and weak leaked signals (marked by red arrows), indicating its deficiencies for complex data. For another, swin-transformer (Fig. 9(c)) also represents limited effects in recovering the consecutive missing traces, resulting energy loss in the restored signals. Overall, the SIT in Fig. 9(d) can accurately reconstruct the missing events, highlighting its inherent advantages for effectively reconstructing DAS data with a big gap. On this basis, we also conduct a detailed analysis by zooming an area of interest in Fig. 9(a) (indicated by yellow dashed rectangle), and the corresponding enlargements are presented in Fig. 10. By analyzing these figures, a similar conclusion can be drawn: Firstly, all three methods demonstrate the capability of DL frameworks to reconstruct consecutive missing traces in DAS seismic data. Second, discontinuous events still can be observed in the reconstruction results of U-Net (Fig. 10(a)) and swin-transformer (Fig. 10(b)). Finally, SIT (Fig. 10(c)) is a competent approach to reconstruct consecutive missing traces in DAS seismic data, even contaminated by intense background noise.

Furthermore, to assess the generalization capability of the proposed methods, we select another field DAS-VSP record (Fig. 11(a)) and employ the SIT and other two frameworks to process it, Fig. 11(b)-(d) gives the corresponding reconstruction results. All DL-based frameworks can achieve effective reconstruction of missing data with consecutive 50 traces. Meanwhile, SIT (Fig. 11(d)) shows the most impressive capability in both noise attenuation and missing trace reconstruction, demonstrating its effectiveness and generalization ability. Therefore, we can find that SIT is effective and may has further application in DAS seismic data processing.

## V. Discussion

### A. Ablation Experiments

In this section, we use ablation experiments to investigate the contributions of key components in SIT, including ET, DT, and UST blocks. Additionally, the effectiveness of the $L_1\_SSIM$ loss function is also discussed.

*1) Network Architecture*

To analyze the effectiveness of ET, DT, and UST in SIT, we conduct following experiments by replacing the target network component with other conventional transformer structures. For a fair comparison, UST is replaced by five transformer blocks, while ET and DT are substitute by two transformer blocks. Meanwhile, the modified networks are trained by the same hyper-parameter settings and training dataset. Subsequently, the obtained reconstruction models are employed to process the incomplete synthetic record depicted in Fig. 4(c). The comparison results of these models are presented in Table III. Notably, it can be found that SIT has a Q value close to 70dB, indicating its advantages over other modified architectures. Overall, we can conclude that all the network components contributes to the promising performance of SIT. In other words, the long-range feature extraction provide by ET and DT and the multi-resolution information offered by UST have a positive effect to improve the reconstruction capability.

*2) Loss Function*

To investigate the effectiveness of $L_1\_SSIM$ loss function other two loss function, including $L_1$ and $L_2$, are used to train SIT. Also the incomplete synthetic record depicted in Fig. 4(c) is selected as the analyzed data. The reconstruction results of different loss function are presented in Table III. The model trained by the $L_1\_SSIM$ loss function achieves the most remarkable performance with a Q value approximately 10dB higher than those of the competing methods. These experimental findings demonstrate the rationality of employing the $L_1\_SSIM$ loss function for the seismic data reconstruction task, also providing a valuable reference for future research.

### B. Analysis of Computational Costs

Computational efficiency is a crucial factor that always needs to be taken into account in DL-based approaches, particularly for massive seismic data processing. In this study, we investigate the computational costs of different methods. Here, the training time and processing time are used to evaluate the computational cost, while Q values are utilized to reflect the performance of different frameworks. By observing the results in Table IV, we can find that SIT has the most heavy computational burden, indicating by its nonnegotiable training time over 56 hours. However, the processing time of SIT is relatively short, only 2.23 seconds. Considering the generalization ability of SIT, the training cost can be shared with the seismic data with similar properties, maybe a few trained models can meet the data requirement in a survey area. From this view point, the computational burden of SIT is acceptable, and the situation may be eased with the development of high efficiency computing equipment. Another important issue is the performance of SIT, which can provide reconstruction results with improved Q value over 60dB. It means SIT is a promising method with a trade-off between computational cost and high reconstruction accuracy.

TABLE III.
COMPARISON OF COMPUTATION COST AND IMPROVED Q VALUE

| Method | U-Net | Swin-transformer | SIT |
|---|---|---|---|
| Training time | 10h32min | 50h05min | 56h34min |
| Processing time | 1.65s | 1.97s | 2.23s |
| Improved Q | 24.60dB | 45.25dB | 68.55dB |

TABLE IV.
ABLATION STUDY ON THE PROPOSED MODEL AND LOSS FUNCTION

| Component usage | The Q value of the results (dB) |
|---|---|
| ET,DT(✗) UST(✓) | 65.59 |
| ET,DT(✓) UST(✗) | 64.34 |
| ET,DT(✓) UST(✓) | 69.30 |
| $L_1$(✓) $L_2$(✗) $L_1\_SSIM$(✓) | 62.07 |
| $L_1$(✗) $L_2$(✓) $L_1\_SSIM$(✓) | 56.83 |
| $L_1$(✗) $L_2$(✗) $L_1\_SSIM$(✓) | 69.30 |

## VI. Conclusion

In this study, we propose a transformer-based architecture, named SIT, for the challenging reconstruction task of consecutive missing traces in DAS-VSP seismic data. Generally, SIT integrates effective network components, including ET, DT, and UST, to extract long-range and multi-resolution features, which may significantly improve the reconstruction accuracy. Meanwhile, we also propose a novel loss function, called $L_1\_SSIM$, to further optimize the training process and enhance the performance of the reconstruction models. Additionally, a high-quality training dataset is also constructed to meet the training requirement and help SIT to learn primary features, which is useful for both synthetic and field DAS data processing. Experimental results have demonstrated that SIT can reconstruct the incomplete DAS data suffering from consecutive missing traces, even in the presence of complex background noise. The quantitative comparison results further validate the effectiveness of SIT, showing superior performance over other popular DL frameworks, including U-Net and swin-transformer. In summary, SIT is a promising framework to cope with the reconstruction task for complex DAS-VSP data, and also exhibits application prospects for massive seismic data processing.

APPENDIX

The concept of local similarity is initially introduced by Fomel as an approach to quantify the resemblance between two different records $p$ and $r$. The calculation of local similarity $s$ between $p$ and $r$ can be performed in the following equation:

$$s = \sqrt{s_1^T s_2} \quad \text{(A-1)}$$

where $s_1$ and $s_2$ represent the solution of the following two least-squares optimization problems:

$$s_1 = arg \min_{s_1} \|p - Rs_1\|_2^2 \quad \text{(A-2)}$$

$$s_2 = arg \min_{s_2} \|r - Ps_2\|_2^2 \quad \text{(A-3)}$$

where P and R represent the diagonal operators of $p$ and $r$, denote as elements of p: $P = \text{diag}(p)$ and R denotes a diagonal operator composed of r: $R = \text{diag}(r)$. Therefore, the least-squares problems in equation A-2 and A-3 can be effectively solved by employing shaping regularization with a smoothness constraint.

$$s_1 = [\lambda_1^2 I + \Gamma(R^T R - )\lambda_1^2 I]^{-1} \Gamma R^T p \quad \text{(A-4)}$$

$$s_2 = [\lambda_2^2 I + \Gamma(P^T P - )\lambda_1^2 I]^{-1} \Gamma P^T r \quad \text{(A-5)}$$

where $\Gamma$ is the smoothing operator, $\lambda_1$ and $\lambda_2$ represent the parameters to control the physical dimensionality and facilitate rapid convergence during the iterative inversion. These two parameters can be selected as $\lambda_1 = \|R^T R\|_2$ and $\lambda_2 = \|P^T P\|_2$.